\documentclass[aps,prx,superscriptaddress,reprint]{revtex4-1}

\usepackage{graphicx}
\usepackage{amsmath}
\usepackage{amsfonts}
\usepackage{amssymb}
\usepackage{bbm}
\usepackage[]{units}
\usepackage{hyperref}
\usepackage{braket}
\usepackage{dsfont}

\newcommand{\nnb}{\nonumber \\}

\newcommand{\bv}{\left( \begin{array}{c}}
\newcommand{\ev}{\end{array} \right)}
\newcommand{\E}{\mathrm{e}}

\newcommand{\st}[1]{_{\text{#1}}}

\newcommand{\I}{\mathrm{i}}

\begin{document}
\title{High-fidelity quantum gates in Si/SiGe double quantum dots}
\author{Maximilian Russ}
\affiliation{Department of Physics, University of Konstanz, D-78457 Konstanz, Germany}
\author{D. M. Zajac} 
\affiliation{Department of Physics, Princeton University, Princeton, New Jersey 08544, USA}
\author{A. J. Sigillito}
\affiliation{Department of Physics, Princeton University, Princeton, New Jersey 08544, USA} 
\author{F. Borjans}
\affiliation{Department of Physics, Princeton University, Princeton, New Jersey 08544, USA} 
\author{J. M. Taylor}
\affiliation{Joint Quantum Institute and Joint Center for Quantum Information and Computer Science, NIST and University of Maryland, College Park, Maryland 20742, USA}
\author{ J. R. Petta}
\affiliation{Department of Physics, Princeton University, Princeton, New Jersey 08544, USA}
\author{Guido Burkard}
\affiliation{Department of Physics, University of Konstanz, D-78457 Konstanz, Germany}
\begin{abstract}
Motivated by recent experiments of ~\citet{Zajac2017}, we theoretically describe high-fidelity two-qubit gates using the exchange interaction between the spins in neighboring quantum dots subject to a magnetic field gradient. We use a combination of analytical calculations and numerical simulations to provide the optimal pulse sequences and parameter settings for the gate operation. We present a novel synchronization method which avoids detrimental spin flips during the gate operation and provide details about phase mismatches accumulated during the two-qubit gates which occur due to residual exchange interaction, non-adiabatic pulses, and off-resonant driving. By adjusting the gate times, synchronizing the resonant and off-resonant transitions, and compensating these phase mismatches by phase control, the overall gate fidelity can be increased significantly.
\end{abstract}
\maketitle

\section{Introduction}

Spin qubits~\cite{Loss1998} implemented in silicon quantum dots~\cite{Zwanenburg2013} are a viable candidate for enabling quantum error corrected quantum computation due to their long coherence times~\cite{Steger2012,Tyryshkin2012,Veldhorst2014,Veldhorst2015} and high-fidelity qubit manipulation~\cite{Takeda2016,Zajac2017,Watson2017}. Experiments using isotopically enriched silicon show single-qubit fidelities $F>99.9\%$ \cite{Yoneda2017} thus exceeding the threshold of quantum-error correction~\cite{Lidar2013}. Successful demonstrations of two-qubit gates\cite{Brunner2011,Veldhorst2015,Zajac2017,Watson2017}, however, show fidelities far below the fault-tolerant threshold, therefore being the limiting factor for large-scale quantum computation. Here, based on the state-of-the-art quantum devices\cite{Zajac2016}, we show a way to implement high-speed and high-fidelity two-qubit gates.

\begin{figure}
\begin{center}
\includegraphics[width=0.79\columnwidth]{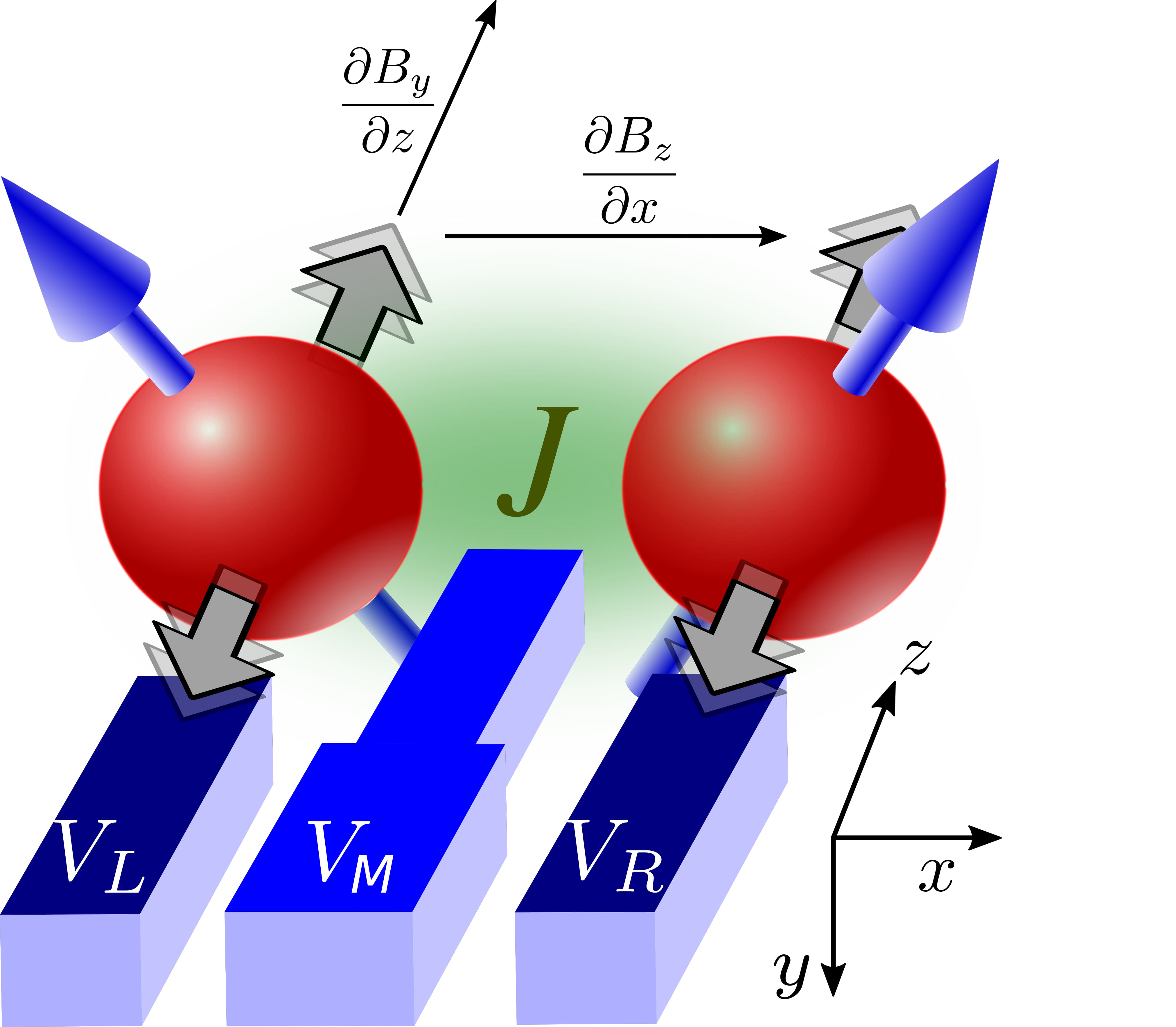}
\caption{Illustration of a gate defined double quantum dot (DQD) occupied with two electron spins inside a large homogenous magnetic field (not shown) and an anisotropic magnetic field $\boldsymbol{B}$ from a micromagnet. A gradient $\frac{\partial B_{z}}{\partial x}$ along the double dot axis ($x$-direction) gives rise to distinguishable spin resonance frequencies. Periodic modulation of the gate voltages ($V_{L}$ and $V_{R}$) shifts the electron position in the $z$-direction (in plane) which in the motion-frame of the electrons ``acts'' as an oscillating magnetic field due to the gradient $\frac{\partial B_{y}}{\partial z}$. An electrostatic barrier gate $V_{M}$ allows for precise control over the exchange interaction $J$ between the spins.}
\label{fig:model}
\end{center}
\end{figure}

High-speed and high-fidelity single-qubit gate operations are achieved using electric dipole spin resonance (EDSR) by shifting the electron position in a slanting magnetic field through the modulation of the electrostatic gate voltages~\cite{Pioro2008,Brunner2011,Otsuka2016,Yoneda2017,Zajac2017,Watson2017}. 
Interconnecting multiple spin qubits is possible through the exchange interaction between electrons in adjacent quantum dots~\cite{Loss1998,Burkard1999,DasSarma2011}. However, the fidelity of these gates is strongly limited by charge noise, which is induced by electric fluctuations of the system, and gives rise to substantial gate operation errors~\cite{Petta2005,Brunner2011}. Higher fidelities can be achieved if the system is operated at a symmetric operation point or sweet spot, where the exchange coupling is first-order insensitive to these fluctuations~\cite{Bertrand2015,Martins2016,Reed2016,Zhang2017,Yang2017}. Alternatively, combining exchange with a strong magnetic field gradient between the electron spins in the dot~\cite{Zajac2017,Watson2017} suppresses the dominating dephasing processes through the large energy splitting of the two-qubit states~\cite{Nichol2017}. Two explicit implementations for two-qubit gates have been successfully demonstrated, an ac pulsed frequency-selective CNOT gate~\cite{Zajac2017,Watson2017} and a dc pulsed CPHASE gate~\cite{Burkard1999b,Meunier2011,Watson2017}. These realizations are still not perfect, both acquiring local phases on the individual spin during the gate-operation due to unitary and non-unitary effects, e.g. charge noise, which have to be identified and compensated. The reduction of the overall gate fidelity due to off-resonant driving still remains an issue without the use of complex pulse shaping~\cite{Economou2015,Ku2017}. 

In this paper, we propose to implement high-fidelity dc CPHASE gates by adding an echo pulse and ac pulsed frequency-selective CNOT gate by synchronizing the resonant and off-resonant Rabi frequencies. We also identify local phases that the individual spins acquire during the CPHASE and CNOT operation due to the influence of the exchange interaction and the resonant and off-resonant driving. The paper is structured as follows. In section~\ref{sec:theomodel}, we begin with the theoretical description of our system. Subsequently, we model the dc pulsed CPHASE gate (section~\ref{sec:dcgate}) and present a high-fidelity implementation in subsection~\ref{ssec:DChigh}. Then, we describe the ac pulsed frequency-selective CNOT gate (section~\ref{sec:ACgate}), provide a synchronized high-fidelity implementation in subsection~\ref{ssec:highfid}, and show its performance under the influence of charge noise in subsection~\ref{ssec:ACnoise}. In section~\ref{sec:con}, we conclude our paper with a summary and an outlook.

\section{Theoretical model}
\label{sec:theomodel}

Our theoretical investigation is inspired by the experiments of Ref.~\cite{Zajac2017}, therefore, we use the same terminology for the theoretical description. The setup (see Fig.~\ref{fig:model}) consists of two gate  defined quantum dots in a Si/SiGe heterostructure operated in the (1,1) regime where $(n_{L},n_{R})$ is defined as the charge configuration with $n_{L}$ electrons in the left and $n_{R}$ electrons in the right dot. A middle barrier gate is biased with voltage $V_{M}$ to tune the exchange interaction~$J$ between the two spins. For our theoretical description we use the Heisenberg Hamiltonian of two neighboring spins that are placed in an inhomogeneous magnetic field
\begin{align}
	H=J(t)(\boldsymbol{S}_{L}\cdot\boldsymbol{S}_{R}-1/4)+\boldsymbol{S}_{L}\cdot\boldsymbol{B}_{L}+\boldsymbol{S}_{R}\cdot\boldsymbol{B}_{R}.
	\label{eq:Ham1}
\end{align}
Here $J$ describes the Heisenberg exchange interaction between the spin in the left dot, $\boldsymbol{S}_{L}$, and the spin in the right dot, $\boldsymbol{S}_{R}$, resulting from the hybridization of the singlet electron wave-function with additional charge states, (2,0) and (0,2). In the Hubbard limit, in the (1,1) charge configuration, exchange is given by $J= 2t_{M}^{2}(U_{L}+U_{R})/[(U_{L}+\varepsilon)(U_{R}-\varepsilon)]$ where $t_{M}=t_{M}(V_{M})$ is the tunneling matrix element between the electron spins which depends on the middle barrier voltage $V_{M}$, $\varepsilon=(V_{L}-V_{R})/2$ is the single-particle detuning between the energy levels of the two spins set by $V_{L}$ and $V_{R}$, and $U_{L}$ and $U_{R}$ are the respective charging energies in the dots. Either biasing the DQD, thus, changing $\varepsilon$, or barrier control, changing $V_{M}$, yields control over the exchange interaction with barrier control being superior if operated at a charge noise sweet spot~\cite{Levy2002,Reed2016,Martins2016} near the center regime of the (1,1) charge state.

The remaining terms in the Hamiltonian~\eqref{eq:Ham1} describe the interaction between the spin and the magnetic field (in energy units) $\boldsymbol{B}_{L}=(0,B^{L}_{y}(t),B^{h}_{z}+B^{L}_{z})^{T}$ and $\boldsymbol{B}_{R}=(0,B^{R}_{y}(t),B^{h}_{z}+B^{R}_{z})^{T}$. The field consists of the homogeneous component $B^{h}_{z}$ which lifts the spin degeneracy, and a spatially dependent field from the micromagnet  $B^{Q}_{z}$ that leads to distinct ESR resonance frequencies for the left and right spin allowing one to individually address each spin. A transverse time-dependent field
\begin{align}
B_{y}^{Q}(t)=B_{y}^{Q,0}+B_{y}^{Q,1}\cos(\omega t+\theta)
\end{align}
occurs from the shift of the electron position in the slanting magnetic field along $x$-direction ($Q=L,R$). This last contribution is further composed of a small static part, $B_{y}^{Q,0}\sim 0$ and a dynamic coupling term, $B_{y}^{Q,1}$ due to an electrostatic modulation of the electrostatic gates, $V_{L}$ and $V_{R}$, with frequency $\omega$.

\begin{figure*}
\begin{center}
\includegraphics[width=1.8\columnwidth]{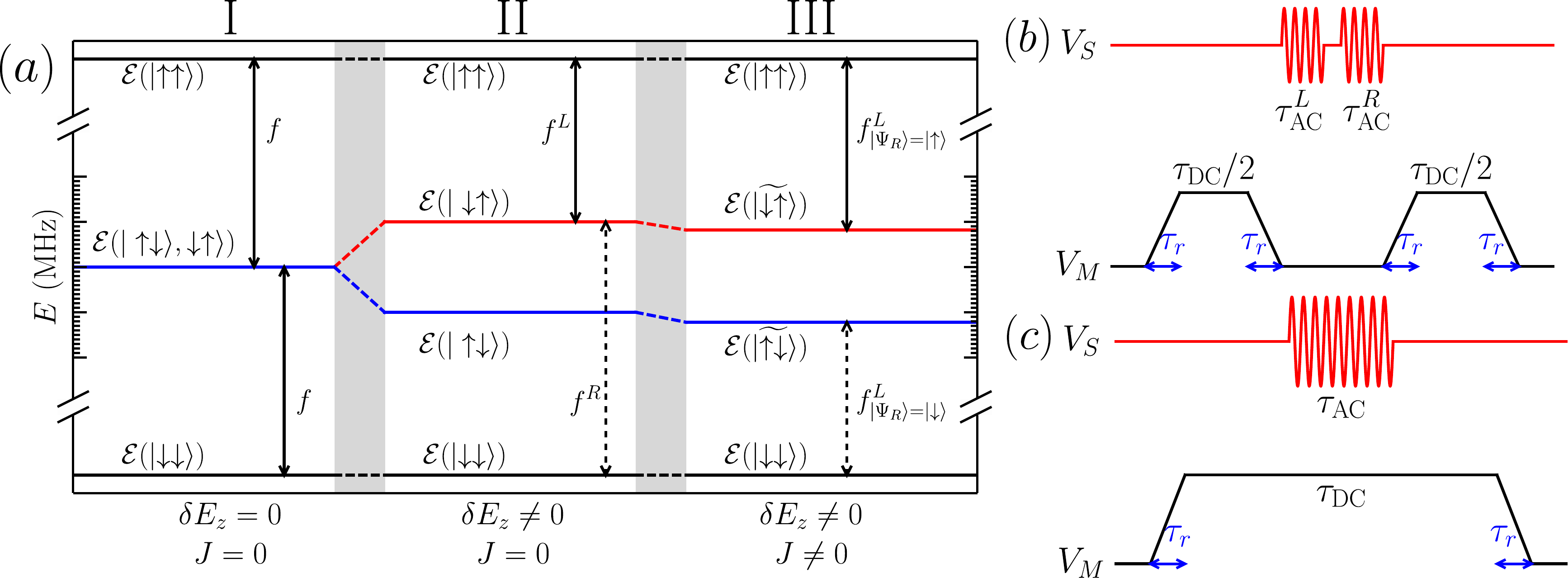}
\caption{(a) Eigenenergies $\mathcal{E}$ of two spins in a double quantum dot in the presence of a magnetic field gradient and relevant transitions between them for three distinct realistic~\cite{Zajac2017} parameter regimes. Regime I describes the case without a magnetic field gradient in which all single-spin flip transitions are energetically degenerate. In the presence of a field gradient (Regime II) the degeneracy of the transition frequencies between the left (solid-line arrow) and right spin (dashed-line arrow) is lifted, making it possible to energetically distinguish between single-qubit operations on the left and right spin. Turning on the exchange interaction, $J\ll\delta E_{z}$, (Regime III) further distinguishes conditional spin-flips from each other, i.e. the left ESR resonance frequency now depends on the right spin state, thus, allowing for frequency-selective entangling two-qubit gates. Conditional spin flip operations are possible: e.g., resonantly driving the transition $\ket{\uparrow\uparrow}\leftrightarrow\ket{\widetilde{\downarrow\uparrow}}$ yields a single-shot CNOT gate. Since the eigenstates $\mathcal{E}(\ket{\widetilde{\downarrow\uparrow}}$) and $\mathcal{E}(\ket{\widetilde{\uparrow\downarrow}}$) are both energetically lowered (by $J/2$ for $J\ll\delta E_{z}$) the transition $\ket{\downarrow\downarrow}\leftrightarrow\ket{\widetilde{\uparrow\downarrow}}$ is detuned from resonance by $\sim J$. 
(b) Schematic pulse sequence of the spin-echo CPHASE gate described in subsection~\ref{ssec:DChigh}. Two adiabatic middle barrier pulses of length $\tau\st{DC}/2$ and ramp length $\tau_{r}$ are combined with two ac pulses of length $\tau\st{AC}^{L,R}$, performing a spin-flip on the left and right spin. 
(c) Schematic pulse sequence of the frequency selective CNOT gate. An adiabatic middle barrier pulse with length $\tau\st{DC}$ and ramp length $\tau_{r}$ is combined with an ac pulse of length $\tau\st{AC}$ resulting in a conditional spin-flip.}
\label{fig:Eig}
\end{center}
\end{figure*}

Addressing each spin as a separate qubit, the Hamiltonian~\eqref{eq:Ham1} can be written in the two-qubit basis $\lbrace \ket{\uparrow\uparrow},\ket{\uparrow\downarrow},\ket{\downarrow\uparrow},\ket{\downarrow\downarrow}\rbrace$ as follows;
\begin{align}
H(t)=\left(
\begin{matrix}
E_{z} & -  \I B^{R}_{y}(t) & -  \I B^{L}_{y}(t)& 0 \\
  \I B^{R}_{y}(t)&-(\delta E_{z}+J)/2& J/2 & - \I B^{L}_{y}(t)\\
  \I B^{L}_{y}(t)& J/2 & (\delta E_{z}-J)/2& - \I B^{R}_{y}(t) \\
 0 &  \I B^{L}_{y}(t) &  \I B_{y}^{R}(t) & -E_{z}
\end{matrix}
\right).
\label{eq:Ham2}
\end{align}
Here we introduced the definitions of the magnetic field difference of strength $\delta E_{z}=B^{R}_{z}-B^{L}_{z}$ and the average Zeeman splitting $E_{z}=B^{h}_{z}+(B^{L}_{z}+B^{R}_{z})/2$. In the absence of exchange, $J\approx 0$, single qubit operations are possible by matching $\omega$ with the resonance frequency, $B_{z}^{h}+B_{z}^{L}$ ($B_{z}^{h}+B_{z}^{R}$), of the left (right) dot separated from each other by $\delta E_{z}$. This corresponds to regime II in Fig.~\ref{fig:Eig}. A large $\delta E_{z}$ is beneficial since it largely separates both resonances in energy allowing for stronger driving, thus, faster gate operations due to the linear dependence of the Rabi frequency $\Omega$ on the modulation strength $B_{y}^{Q,1}$.

\section{DC Entangling gates: Strong and weak exchange}
\label{sec:dcgate}

Two-qubit gates between neighboring single-spin qubits are realizable using the exchange interaction between the spins~\cite{Loss1998,Petta2005} with or without a magnetic field gradient~\cite{Burkard1999b,Meunier2011}. If the exchange energy dominates the Hamiltonian~\eqref{eq:Ham2}, i.e. $J\gg\delta E_{z}$, the (approximate) two-qubit eigenstates are the spin singlet, $\ket{\uparrow\downarrow}-\ket{\downarrow\uparrow}$ and triplets, $\ket{\uparrow\uparrow}$, $\ket{\uparrow\downarrow}+\ket{\downarrow\uparrow}$, $\ket{\downarrow\downarrow}$, and the resulting operation yields (for $\delta E_{z}=0$) the entangling $\sqrt{\text{SWAP}}$-gate. Sequential implementation of two $\sqrt{\text{SWAP}}$-gates and single-qubit rotations yields a CNOT-gate~\cite{Loss1998}. In the case of weak exchange, i.e. $J\ll\delta E_{z}$, the two-qubit states are effectively the product states $\ket{\uparrow\uparrow}$, $\ket{\widetilde{\uparrow\downarrow}}$, $\ket{\widetilde{\downarrow\uparrow}}$, $\ket{\downarrow\downarrow}$ with small corrections in $\ket{\widetilde{\uparrow\downarrow}}$ and $\ket{\widetilde{\downarrow\uparrow}}$ due to spin-charge hybridization. In this limit the exchange interaction  yields a conditional phase (CPHASE) gate. In this paper we focus on the regime $J\ll\delta E_{z}$ which is typical for DQD systems in the presence of a micromagnet. However, for adiabatic pulses, with ramp time $\tau_{r}\gg J/\delta E_{z}^{2}$, both implementations are equivalent. Note, that the criteria for the adiabatic regime is usually fulfilled in state-of-the-art devices\cite{Zajac2017,Yoneda2017,Watson2017}. For an adiabatic pulse the instantaneous eigenvalues of the Hamiltonian~\eqref{eq:Ham2} are given as follows;
\begin{align}
\mathcal{E}(\ket{\uparrow\uparrow})&=E_{z},\label{eq:eigenDC1}\\
\mathcal{E}(\ket{\widetilde{\uparrow\downarrow}})&=\left(-J-\sqrt{J^{2}+\delta E_{z}^2}\right)/2,\label{eq:eigenDC2}\\
\mathcal{E}(\ket{\widetilde{\downarrow\uparrow}})&=\left(-J+\sqrt{J^{2}+\delta E_{z}^2}\right)/2,\label{eq:eigenDC3}\\
\mathcal{E}(\ket{\downarrow\downarrow})&=-E_{z}.
\label{eq:eigenDC4}
\end{align}
Figure~\ref{fig:Eig}~(a) shows the eigenenergies for three different parameter regimes for $J\ll\delta E_{z}$. Note, that for $J\ll\delta E_{z}$ one can use the expansion $\sqrt{J^{2}+\delta E_{z}^2}\approx\delta E_{z}+\frac{J^{2}}{2\delta E_{z}}$ simplifying the  expressions in Eqs.~\eqref{eq:eigenDC2}~and~\eqref{eq:eigenDC3}. The time evolution of an adiabatic exchange pulse of length $\tau$ in the rotating frame $\widetilde{H}(t)=R^{\dagger}HR+\I \dot{R}^{\dagger}R$ with $R=\exp\left[-\I \omega t(S_{z,L}+S_{z,R})/\hbar\right]$ and $\hbar\omega=E_{z}/2$ is given by 
\begin{align}
U(\tau)=\text{diag}\left[1,\,\,\E^{\I\left(J+\sqrt{J^{2}+\delta E_{z}^2}\right)\frac{\tau}{2\hbar}},\,\,\E^{\I\left(J-\sqrt{J^{2}+\delta E_{z}^2}\right)\frac{\tau}{2\hbar}},1\right].
\end{align}
Note that here $g(J)\tau$, for arbitrary functions $g$, is defined as $g(J)\tau\equiv\int_{0}^{\tau}g(J(t))dt$.
The time evolution up to a global phase can be decomposed into two parts as $U=U\st{ent}U\st{loc}$ with an entangling term 
\begin{align}
U(\tau)\st{ent}=\exp\left(\I J \tau\, S_{z,L}S_{z,R}/\hbar\right)
\label{eq:UDCent}
\end{align}
and an accumulated local phase
\begin{align}
U(\tau)\st{loc}=&\exp\left(-\I \sqrt{J^{2}+\delta E_{z}^2}  (S_{z,L}-S_{z,R})\tau/\hbar\right)\nnb
\approx& \exp\left[-\I \left(\delta E_{z}+\frac{J^{2}}{2\delta E_{z}}\right) (S_{z,L}-S_{z,R})\tau/\hbar\right].
\label{eq:UDCphase}
\end{align}
For gate times being odd integer multiples of $\hbar \pi/J$, $\tau=(2n+1)\hbar\pi/J$ and the time-evolution~\eqref{eq:UDCent} is equivalent to CPHASE up to single-qubit $z$-rotations~\cite{Meunier2011}. Even multiples, $\tau_{\text{DC}}\equiv\tau=2\hbar\pi n/J$, correspondingly yield identity up to local $S_{z}$-rotations $\E^{\I\Phi_{L}S_{z,L}/2}$ and $\E^{\I\Phi_{R}S_{z,R}/2}$ for the left and right spin. From Eq.~\eqref{eq:UDCphase} we find the following expressions for the phases,
\begin{align}
\Phi_{L}^{\text{DC}}&=-2\pi n\sqrt{J^{2}+\delta E_{z}^2}/J \label{eq:DCphaseL},\\
\Phi_{R}^{\text{DC}}&=2\pi n\sqrt{J^{2}+\delta E_{z}^2}/J \label{eq:DCphaseR}.
\end{align}
The correction of these phases will be discussed in subsection~\ref{ssec:highfid}. The fact that the dc CPHASE operation can be cancelled out will be important for the ac gate discussed in section~\ref{sec:ACgate}.

\subsection{High fidelity dc implementation}
\label{ssec:DChigh}

In the experimental configuration described above, the magnetic gradient can exceed the nominal exchange splitting due to residual tunneling to a high degree. Thus we can examine the type of gates that may provide high fidelity operation in that environment. We assume that in the case of a nominal zero induced exchange, the two spin system is described by a Hamiltonian
\begin{align}
H_0 = E_z (S_{z,L} + S_{z,R}) + \delta E_z (S_{z,L} - S_{z,R})/2
\end{align}
where we used the same definitions as above. We have neglected the transverse magnetic gradient which causes individual spin flips and is suppressed by the Zeeman energy $E_z$. We also neglect the induced double-spin flip transitions by the diagonalization of the physical Hamiltonian under finite residual exchange $J_0$.  

We consider the inclusion of an echo mechanism for removing the excess phase terms in the exchange gate as well as potential unknown magnetic gradients. In order to examine the spin parity subspaces efficiently, we define new Pauli matrices, $\sigma_x = S_{+,L} S_{-,R} + {H.c.}$, $\sigma_z = S_{z,L} - S_{z,R}$, and the projector $P = (1 - 4 S_{z,L} S_{z,R})/2$, acting only in the odd parity space. In a similar manner, we define $\tau_z = S_{z,L} + S_{z,R}$ and $\tau_x = S_{+,L} S_{+,R} + S_{-,L} S_{-,R}$. In this basis, the (time-dependent) Hamiltonian~\eqref{eq:Ham1} neglecting transverse gradient fields is 
\begin{align}
H(t) = E_{z} \tau_z + \delta E_z \sigma_z/2 - J(t) [P - \sigma_x]/2
\end{align}
with $J(t)$ being the ramp-induced exchange. We also note that a $\pi$ pulse on both spins about the $x$-axis corresponds, up to a global phase, to the unitary $\tau_x + \sigma_x$ while a pair of $y$-axis pulses would correspond to $\tau_x  - \sigma_x$.

For both fast and slow exchange pulses of length $T$, we see that the even parity space only undergoes evolution according to $E_{z} \tau_z$, and thus will effectively factor out after inclusion of the $\pi$ pulses shown in Fig.~\ref{fig:Eig}~(b). Meanwhile, the odd parity space undergoes nontrivial evolution, due to both the overall phase evolution $-\int_0^T J dt/2$ and from the rotations in the subspace about the axis $(J,0,\delta E_{z})$.

We first consider fast instantaneous changes of $J$. The time evolution is then stroboscopically given by rotations for a controlled period about various axes in the odd parity ($P$) subspace. A simple exchange pulse corresponds to
\begin{align}
U_J =& (\mathbb{I}-P) \E^{-\I T E_{z} \tau_z} +\\
&\ \ P \E^{\I \Phi_x} \left[\cos(\Omega_J T) - \I \frac{J \sigma_x + \delta E_z \sigma_z}{2\Omega_J} \sin(\Omega_J T) \right] \nonumber
\end{align}
with $\Omega_J = \sqrt{\delta E_z^2 + J^2}/2$ and $\Phi_x = \int_0^T J dt/2 = J T/2$. We note that for $\Omega_J T = n \pi$ with integer $n$, we obtain a CPHASE gate with phase $\Phi_x = \frac{J}{\sqrt{\delta E_z^2 + J^2}} n \pi$, as the $\sin$ terms vanishes, which presents one way to remove the excess phase. 

We now add the additional rotations of individual spins in the middle of the sequence. Towards this end, we would like to understand how this gate behaves in the rotating frame in which we apply our single qubit gates. Note that in this subsection, our rotating frame has a different frequency for each of the two spins, representing the frequencies of the local oscillators used to drive individual spin resonance (motivated by experiment~\cite{Zajac2017}). If we envision starting $U_J$ at time $t_1$ and ending it at time $t_2 = t_1 + T$, we need to know the rotating frame state at the end of the sequence.  We can move $U_J$ to this rotating frame, $R=\E^{-\I (\omega_1 S_{z,L} + \omega_2 S_{z,R}) t}$, defined by the qubit frequencies $\omega_1 = (E_z + \delta E_z/2)/\hbar$ and $\omega_2 = (E_z - \delta E_z/2)/\hbar$ and applying the unitary transformation $\ket{\psi_{\text{rf}}(t)} = R^{\dagger} \ket{\psi_{\text{lab}}(t)}$.  Thus, we have
\begin{align}
\ket{\psi_{\text{rf}}(t_2)} = R^\dagger U_JR \ket{\psi_{\text{rf}}(t_1)} 
= U_J^{\text{rf}}(t_2,t_1) \ket{\psi_{\text{rf}}(t_1)},
\end{align}
where we move back to the lab frame, apply $U_J$, then move back to the rotating frame.

In total, in the rotating frame, we find
\begin{widetext}
\begin{align}
U_J^{\text{rf}}(t_1+T,t_1)  
&= (\mathbb{I}-P) + P \E^{\I \Phi_x} \E^{\I \delta E_z T \sigma_z/4} \left[\cos(\Omega_J T) - \I \frac{J \sigma_{\mu}  + \delta E_z \sigma_z}{2\Omega_J} \sin(\Omega_J T) \right] \E^{\I \delta E_z T \sigma_z/4}
\label{eq:fastRF}
\end{align}
\end{widetext}
where $\sigma_\mu = \cos(\delta E_z (2t_1+T)/2) \sigma_x + \sin(\delta E_z (2 t_1 + T)/2) \sigma_y$. Thus the wait time $t_1$ enters in the definition of the rotation axis. While we do not necessarily want any such rotation, as the diagonal term $\sim J$ will perform a CPHASE-like evolution, we will have to be careful to make certain effects from this are removed.

One approach for removing the flip-flop effect consists in moving adiabatically with respect to $\delta E_z$. Diagonalization of $H(t)$ in the rotating frame yields $\tilde{H}^{\text{rf}} = - P J(t)/2 + (\Omega_J(t) - \delta E_z/2) \sigma_z$ with $\Omega_J(t) = \sqrt{\delta E_z^2 + J(t)^2}/2$. Small non-adiabatic corrections enter with a $\sigma_+$ term, which behave in a similar manner to the $\sigma_x$ term given by $U_J$ for the fast case. The net result of the adiabatic case in the rotating frame is
\begin{align}
U_{\text{ad}}^{\text{rf}} = (1-P) +
P \E^{\I \Phi_{\text{ad}}}  \E^{-\I \sigma_z \phi_z}
\end{align}
with $\Phi_{\text{ad}} = \int_0^T J(t) dt/2$ and $\phi_z = \int_0^T \sqrt{\delta E_z^2 + J(t)^2} dt/2 - \delta E_z T/2$. Thus we see that the adiabatic $J$ pulse leads to an extra single qubit $z$ rotation for both spins, in addition to the desired CPHASE-like operation. Note, that this phase in the rotating frame of the individual spins is equivalent to the phase given by Eqs.~\eqref{eq:DCphaseL}~and~\eqref{eq:DCphaseR} for $n=1$.

We now consider a more general solution to the extra phase evolution (corresponding to a potentially undesired set of single qubit $z$ rotations) as well as the extra rotation about the $\mu$-axis. Specifically, we consider two $\pi$ pulses about the $x$-axis on the qubits in between two CPHASE-like unitaries (see Fig.~\ref{fig:Eig}~(b)). For the adiabatic case, we have
\begin{align}
U_{c,\text{ad}} &= U^{\text{rf}}_{\text{ad}} (4 S_{x,L} S_{x,R}) U^{\text{rf}}_{\text{ad}}  \\
    &= (4 S_{x,L} S_{x,R}) \left[ (1-P) + P \E^{\I \Phi_{\text{ad}}} \E^{\I \sigma_z \phi_z} \right] \nonumber \\
    &\ \ \times \left[ (1-P) + P \E^{\I \Phi_{\text{ad}}} \E^{-\I \sigma_z \phi_z} \right] \\
    &= (4 S_{x,L} S_{x,R}) \left[(1-P) + P \E^{2 \I \Phi_{\text{ad}}} \right] \\
    &= 4 S_{x,L} S_{x,R} \E^{-2 \I \Phi_{\text{ad}} (S_{z,L} + S_{z,R})} U_{\rm CPHASE},
\end{align}
where $U_{\rm CPHASE} = \textrm{diag}[1,1,1,\E^{-2 \I \Phi_{\text{ad}}}]$. For the special case of $2\Phi_{\text{ad}} = \pi$, we find for our gate
\begin{align}
U_{\pi,\text{ad}} = (4 S_{y,L} S_{y,R}) U_{\text{CZ}}.
\end{align}

Returning to the fast pulse version of the gate, Eq.~\eqref{eq:fastRF}, we see that the same $\pi$ pulses in the middle lead to
\begin{align}
U_{c,\text{fast}} &= U_J^{\text{rf}}(t_2+T,t_2) (4 S_{x,L} S_{x,R}) U_J^{\text{rf}}(T,0) \\
&= U_J^{\text{rf}}(t_2+T,t_2) (\tau_x + \sigma_x) U_J^{\text{rf}}(T,0) \\
&= (\mathbb{I} - P) \tau_x + P \E^{2 \I \Phi_x} \E^{\I \delta E_z (t_2 + T)\sigma_z/2} \left[c - \vec{n} \cdot \vec{\sigma} s \right] \nonumber \\
& \ \ \times \E^{-\I \delta E_z t_2 \sigma_z/2} \sigma_x \E^{\I \delta E_z T \sigma_z/2} \left[c - \vec{n} \cdot \vec{\sigma} s \right] 
\end{align}
where $c = \cos(\Omega_J T)$, $s = \sin(\Omega_J T)$, $\vec{n} = (J,0,\delta E_z)/(2\Omega_J)$.
In order to remove the terms proportional to $\vec{n}$ in the above, we need the action of the intermediate rotation $\E^{-\I \delta E_z t_2 \sigma_z/2} \sigma_x \E^{\I \delta E_z T \sigma_z/2}$ to correspond to $\sigma_y$. That requires $\sin(\delta E_z (t_2 + T)/2) = \pm 1$.

Furthermore, we want the equivalent unitary after the sequence,
\[
\E^{\I \delta E_z/2 (t_2 + T)\sigma_z} \sigma_y
\]
to be $\sigma_x$, which in turn requires $\delta E_z (t_2+T)/2 = (2n+1)\pi$ with integer $n$. Conveniently, these are the same requirement.

\section{Resonant single step CNOT gate}
\label{sec:ACgate}

Additional controllability is given if the adiabatic dc exchange pulse is combined with microwave ac driving, $B_{y}^{Q}(t)\propto\cos(\omega t +\varphi)$, matching the transition frequencies between the two-qubit states which allows for direct conditional spin-flips. The gate sequence is outlined in Fig.~\ref{fig:Eig}~(b) and the basic concept is visualized in Fig.~\ref{fig:Eig}~(a) regime III. With the exchange interaction turned on the energy of both eigenstates $\ket{\widetilde{\uparrow\downarrow}}$ and $\ket{\widetilde{\downarrow\uparrow}}$ is lowered by $\sim J/2$, providing in total six energetically distinct resonance frequencies in the spectrum. There are four entangling transitions corresponding to the four conditional spin-flips. For example, inducing a resonant spin-flip between the states $\ket{\uparrow\uparrow}\leftrightarrow \ket{\widetilde{\downarrow\uparrow}}$ yields a CNOT with the right qubit as control and the left qubit as target gate as the following truth-table shows
\begin{align}
\begin{split}
\ket{\uparrow\uparrow}&\rightarrow \ket{\downarrow\uparrow}\\
\ket{\uparrow\downarrow}&\rightarrow \ket{\uparrow\downarrow}\\
\ket{\downarrow\uparrow}&\rightarrow \ket{\uparrow\uparrow}\\
\ket{\downarrow\downarrow}&\rightarrow \ket{\downarrow\downarrow}.
\end{split}
\label{eq:truthCNOT}
\end{align}
In the remainder of this article, we always refer to this implementation of CNOT, however, in experiments other transitions can be resonantly driven as well, giving access to a much larger set of two-qubit quantum gates.
 
From the eigenenergies~\eqref{eq:eigenDC1}-\eqref{eq:eigenDC4} the corresponding resonance frequencies of the four conditional transitions are given as follows;
\begin{align}
f^{L}_{\ket{\Psi_{R}}=\ket{\downarrow}}\equiv& |\mathcal{E}(\ket{\downarrow\downarrow})-\mathcal{E}(\ket{\widetilde{\uparrow\downarrow}})|\nnb
=&E_{z}+\left(-J-\sqrt{J^{2}+\delta E_{z}^2}\right)/2\label{eq:trans1},\\
f^{L}_{\ket{\Psi_{R}}=\ket{\uparrow}}\equiv& |\mathcal{E}(\ket{\widetilde{\downarrow\uparrow}})-\mathcal{E}(\ket{\uparrow\uparrow})|\nnb
=&E_{z}+\left(J-\sqrt{J^{2}+\delta E_{z}^2}\right)/2\label{eq:trans2},\\
f^{R}_{\ket{\Psi_{L}}=\ket{\downarrow}}\equiv&|\mathcal{E}(\ket{\downarrow\downarrow})-\mathcal{E}(\ket{\widetilde{\downarrow\uparrow}})|\nnb
=&E_{z}+\left(-J+\sqrt{J^{2}+\delta E_{z}^2}\right)/2\label{eq:trans3},\\
f^{R}_{\ket{\Psi_{L}}=\ket{\uparrow}}\equiv& |\mathcal{E}(\ket{\widetilde{\uparrow\downarrow}})-\mathcal{E}(\ket{\uparrow\uparrow})|\nnb
=&E_{z}+\left(J+\sqrt{J^{2}+\delta E_{z}^2}\right)/2.
\label{eq:trans4}
\end{align}
One important observation is that the splitting between the conditional spin-flips is always provided by exchange,
\begin{align}
f^{L}_{\ket{\Psi_{R}}=\ket{\uparrow}}-f^{L}_{\ket{\Psi_{R}}=\ket{\downarrow}}=f^{R}_{\ket{\Psi_{L}}=\ket{\uparrow}}-f^{R}_{\ket{\Psi_{L}}=\ket{\downarrow}}=J.
\label{eq:exMeas}
\end{align}

\subsection{High fidelity ac implementation}
\label{ssec:highfid}

We have shown so far that we can effectively cancel out the CPHASE gate from the dc dynamics of our frequency selective gate by appropriately timing the length of the dc exchange pulse $t\st{DC}=2\pi n/J$ where $n$ is a positive integer. This can be thought of applying CPHASE twice (times $n$) which undo each other. However, there are two additional effects which will disturb the gate if not treated appropriately. The first effect results from the off-resonant driving of nearby transitions, which can be important if $J$ is comparable to the Rabi frequency, and a second effect originates from relative phase accumulation of the spins during the microwave drive. Below we discuss both effects and how they can be avoided.

In the experiment described in Ref.~\cite{Zajac2017} the gates are driven at the resonance frequency $\omega=f^{L}_{\ket{\Psi_{R}}=\ket{\uparrow}}$ during a dc exchange pulse which flips the left spin if and only if the right spin is in the state $\ket{\Psi_{R}}=\ket{\uparrow}$ inducing a transition between the $\ket{\uparrow\uparrow}$ and $\ket{\widetilde{\downarrow\uparrow}}$ states. However, the energy separation of the transition frequency $f^{L}_{\ket{\Psi_{R}}=\ket{\uparrow}}$ and the transition frequency for an opposite right spin, $\ket{\Psi_{R}}=\ket{\downarrow}$, is given by the exchange interaction strength $J$ (see Eq.~\eqref{eq:exMeas}). In the regime of operation ~\cite{Zajac2017} $\delta E_{z}\gg J\approx\unit[20]{MHz}$ the transition between the states $\ket{\uparrow\downarrow}$ and $\ket{\widetilde{\downarrow\downarrow}}$ is also driven and gives rise to off-resonant Rabi dynamics. Other transitions, $f^{R}_{\ket{\Psi_{L}}=\ket{\uparrow}}$ and $f^{R}_{\ket{\Psi_{L}}=\ket{\downarrow}}$, are even further off-resonant because they are separated in energy by $\delta E_{z}$, and will be neglected here.

Starting with the Hamiltonian~\eqref{eq:Ham2} in the rotating frame $\widetilde{H}(t)=R^{\dagger}HR+\I \dot{R}^{\dagger}R$ with $R=\exp\left[-\I \omega t(S_{z,L}+S_{z,R})/\hbar\right]$ and neglecting fast oscillations, we find in the instantaneous adiabatic basis $\lbrace \ket{\uparrow\uparrow},\ket{\widetilde{\downarrow\uparrow}},\ket{\widetilde{\uparrow\downarrow}},\ket{\downarrow\downarrow}\rbrace$ for $J\ll \delta E_{z}$
\begin{widetext}
\begin{align}
\widetilde{H}(t)\approx\frac{1}{2}\left(
\begin{matrix}
2(E_{z}-\omega)  & -  \I \alpha_{1}^{*} & -  \I \beta_{1}^{*} & 0 \\
  \I \alpha_{1} &  -J +(\delta E_{z}+\frac{J^{2}}{2\delta E_{z}})& 0 & - \I \beta_{2}^{*} \\
   \I \beta_{1} & 0 &-J -(\delta E_{z}+\frac{J^{2}}{2\delta E_{z}}) & - \I \alpha_{2}^{*} \\
 0  &  \I \beta_{2} & \I \alpha_{2} & -2(E_{z}-\omega) \\
\end{matrix}
\right).
\label{eq:HamRWA}
\end{align}
\end{widetext}
\begin{figure*}
\begin{center}
\includegraphics[width=0.8\textwidth]{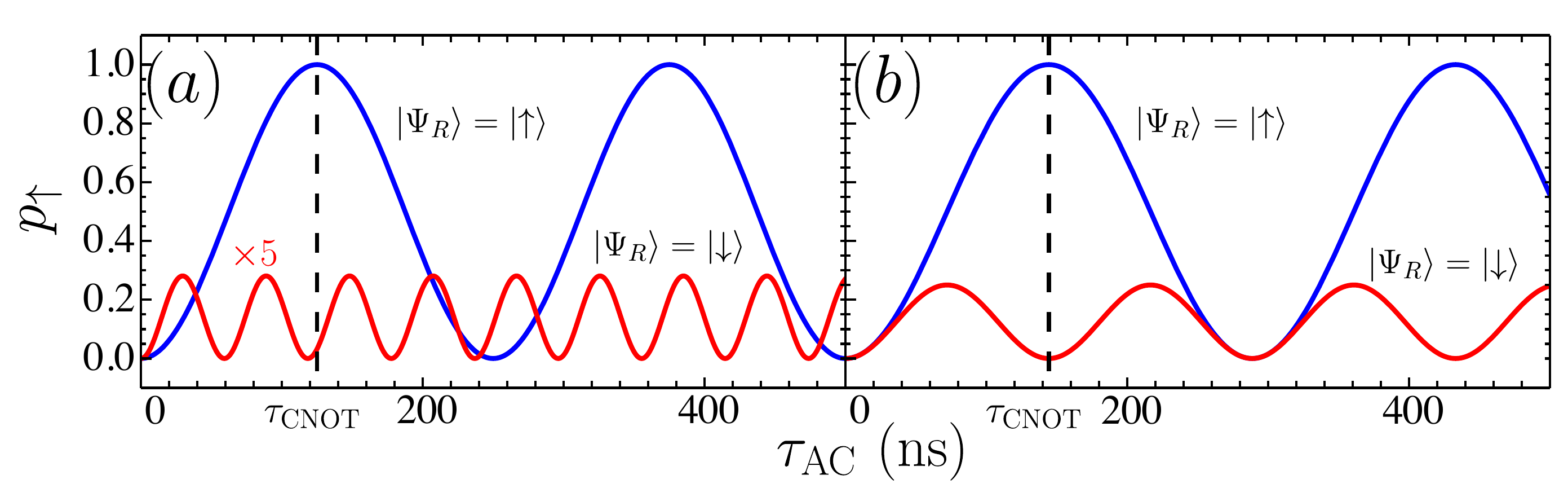}
\caption{Schematic plot of the resonant (blue) and off-resonant (red) transition probabilities with (a) desynchronized and (b) synchronized Rabi frequencies $\Omega$ and $\widetilde{\Omega}$ (see Eq.~\eqref{eq:detRabi}). A CNOT gate is provided at time $\tau\st{CNOT}$ where the resonant driving exactly flips the spin for $\ket{\Psi_{R}}=\ket{\uparrow}$ (blue). (a) Using an arbitrary Rabi frequency $\Omega$, we find that the frequencies are desynchronized and the off-resonant driving also yields a finite population of the flipped spin for $\ket{\Psi_{R}}=\ket{\downarrow}$ (red). (b) Synchronized resonant Rabi frequency $\Omega$ and off-resonant Rabi frequency $\widetilde{\Omega}$ which avoids any undesired population for $\ket{\Psi_{R}}=\ket{\downarrow}$. We choose $m=0$ and $n=1$ for the fastest realization of the synchronized CNOT gate yielding an ac pulse length $\tau\st{AC}\approx\unit[94]{ns}$ using parameters from the experiment in Ref.~\cite{Zajac2017}. Note that to enhance the visibility, we rescaled the off-resonant state probabilities in (a) by a factor of $5$.}
\label{fig:Synch}
\end{center}
\end{figure*}
Here $\alpha_{1,2}\approx [\pm B^{L,1}_{y}+ B^{R,1}_{y} J/(2\delta E_{z})]\,\, \E^{\I \theta}$ and $\beta_{1,2}\approx [\mp B^{R,1}_{y}+ B^{L,1}_{y} J/(2\delta E_{z})]\,\, \E^{\I \theta}$ are the effective microwave driving amplitudes after transforming into the adiabatic basis.
Nearby the resonance frequency $\omega-\delta\omega=f^{L}_{\ket{\Psi_{R}}=\ket{\uparrow}}\approx E_{z}-\lbrace\delta E_{z}+J[1-J/(2\delta E_{z})]\rbrace/2$, $\beta_{1,2}\approx 0$, the Hamiltonian~\eqref{eq:HamRWA} decouples into two blocks, $\lbrace \ket{\uparrow\uparrow},\ket{\widetilde{\downarrow\uparrow}}\rbrace$ and $\lbrace \ket{\widetilde{\uparrow\downarrow}},\ket{\downarrow\downarrow}\rbrace$
which are separated in energy by $\sim\delta E_{z}$ (see Eq.~\eqref{eq:exMeas}) and evolve independently in time. Therefore, we find in the basis $\lbrace \ket{\uparrow\uparrow},\ket{\widetilde{\downarrow\uparrow}},\ket{\widetilde{\uparrow\downarrow}},\ket{\downarrow\downarrow}\rbrace$
\begin{widetext}
\begin{align}
\widetilde{H}(t)\approx\frac{1}{2}\left(
\begin{array}{cc|cc}
-J +(\delta E_{z}+\frac{J^{2}}{2\delta E_{z}}) -\delta\omega & - \I \alpha_{1}^{*}  & 0& 0 \\
 \I \alpha_{1} & -J +(\delta E_{z}+\frac{J^{2}}{2\delta E_{z}})& 0 & 0 \\
 \hline
0  & 0 &-J -(\delta E_{z}+\frac{J^{2}}{2\delta E_{z}}) & - \I \alpha_{2}^{*} \\
0 & 0 &  \I \alpha_{2} & J -(\delta E_{z}+\frac{J^{2}}{2\delta E_{z}}) +\delta\omega \\
\end{array}
\right).
\label{eq:HamRWAblock}
\end{align}
\end{widetext}

\begin{figure}[t]
\begin{center}
\includegraphics[width=1.\columnwidth]{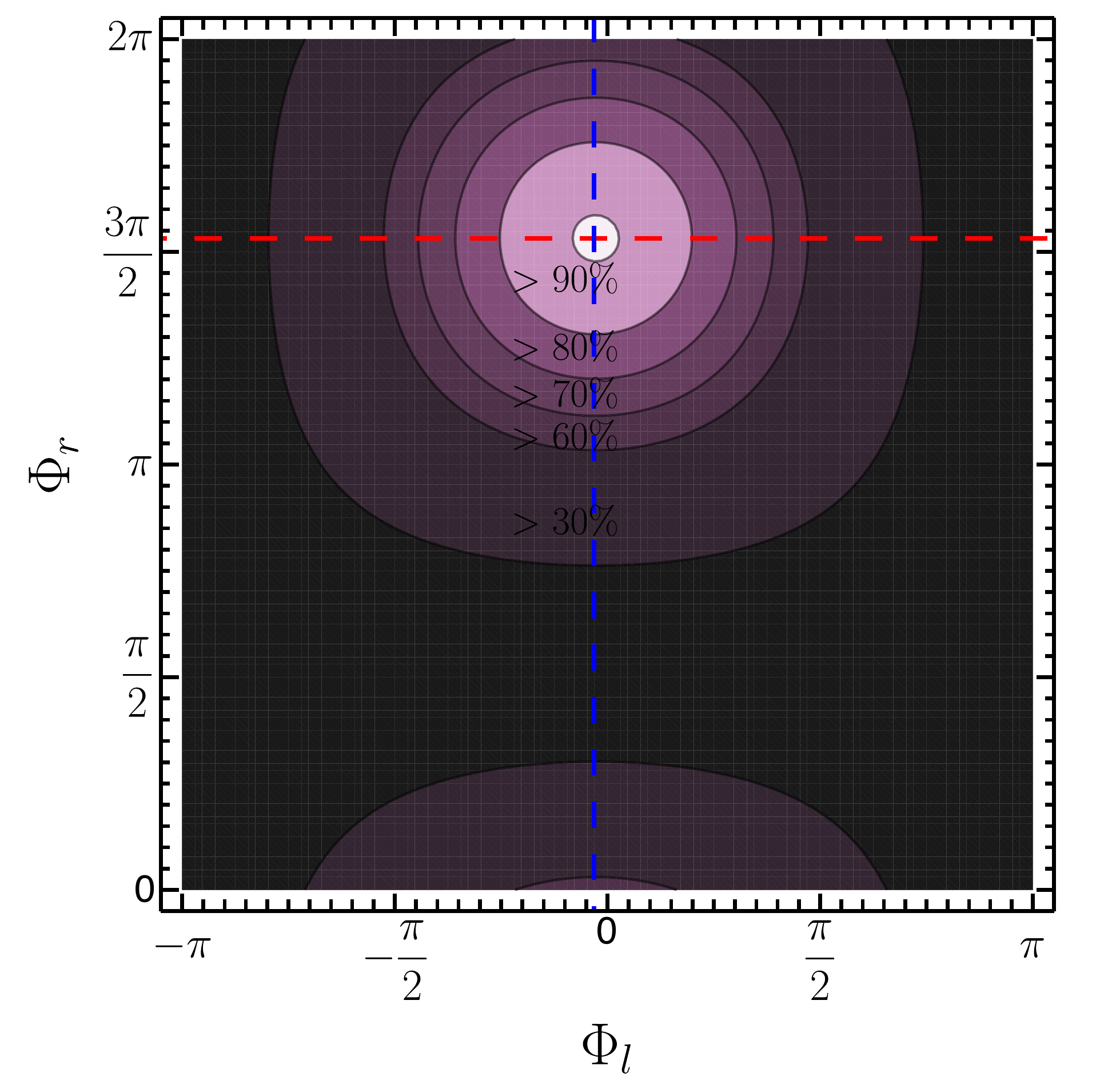}
\caption{Simulated CNOT gate fidelity (in percent) as a function of the compensated phase $\Phi_{L}$ and $\Phi_{R}$ on the left spin (target) and right spin (control) after the ac drive. Compensation is provided for a $z$-Rotation $\hat{Z}(\Phi)$ on the left spin with $\Phi_{l}=2\pi k_{l}- \Phi_{L}$ and on the right spin $\Phi_{r}=2\pi k_{r}- \Phi_{R}$ with integers $k_{l}$ and $k_{r}$, thus $\Phi=(\Phi_{l},\Phi_{r})\approx(-0.03\pi,-0.469\pi)$ (blue and red dashed lines). For the simulation we used the following parameters related to the experiment in Ref.~\cite{Zajac2017}; $J=\unit[19.7]{MHz}$, $\delta E_{z}=\unit[210]{MHz}$, $\tau_{r}=\unit[5]{ns}$, $\tau\st{CNOT}=\unit[94]{ns}$ (corresponding to $\alpha=1/(2\tau\st{CNOT})$, $\tau\st{DC}=\unit[198]{ns}$, and $m=0$ and $n=1$ in Eq.~\ref{eq:detRabi})}.
\label{fig:phasecorrection}
\end{center}
\end{figure}

For $\delta\omega=0$, only $f^{L}_{\ket{\Psi_{R}}=\ket{\uparrow}}$ (top-left block) is resonant and yields full Rabi oscillations with a Rabi frequency $\Omega=|\alpha_{1}|/\hbar$ while $f^{L}_{\ket{\Psi_{R}}=\ket{\downarrow}}$ (bottom-right block) is detuned (off-resonant) by $J$, therefore, performing partial spin-flips with the detuned Rabi frequency $\widetilde{\Omega}=\sqrt{|\alpha_{2}|^{2}+J^{2}}/\hbar$. Since the time evolution of each $2\times 2$ block can be computed individually we find the following time evolutions of each block for $\theta=3\pi/2$,
\begin{align}
U_{\ket{\Psi_{R}}=\ket{\uparrow}}=\E^{-\frac{\I t c_{1}}{2}}&\left[\cos\left(\frac{\Omega t}{2}\right)\mathds{1}+ \I \sin \left(\frac{\Omega t}{2}\right)\sigma_{x} \right]\label{eq:SubUp},\\
U_{\ket{\Psi_{R}}=\ket{\downarrow}}=\E^{-\frac{\I t c_{2}}{2}}&\left[\cos\left(\frac{\widetilde{\Omega} t}{2}\right)\mathds{1}+ \I \sin \left(\frac{\widetilde{\Omega} t}{2}\right)\right.\nnb
&\left.\times\left(\frac{|\alpha_{2}|}{2\,\hbar\, \widetilde{\Omega}}\sigma_{x}-\frac{J}{2\,\hbar\, \widetilde{\Omega}}\sigma_{z}\right) \right].\label{eq:SubDown}
\end{align}
with the frequencies $\hbar c_{1}=-J +(\delta E_{z}+\frac{J^{2}}{2\delta E_{z}})$ and $\hbar c_{2}=-\delta E_{z}-\frac{J^{2}}{2\delta E_{z}}$.
Setting $t= \pi\, (2m+1)/\Omega$ with integer $m$ yields a spin flip in the $\ket{\Psi_{R}}=\ket{\uparrow}$ block. In order to cancel the dynamics of the off-resonant states, we synchronize the Rabi frequencies by setting
\begin{align}
\Omega=\frac{2m+1}{2n}\widetilde{\Omega},
\label{eq:detRabi}
\end{align}
with an integer $n$. This can be achieved by adjusting the ac driving strength $B^{L,1}_{y}$. Considering $B^{L,1}_{y}=B^{R,1}_{y}$, we find the following analytical result for the ac driving strength,
\begin{align}
B^{L,1}_{y}=a_{n,m}\equiv\pm\frac{J}{\sqrt{\frac{4n^{2}}{(2m+1)^{2}}\left(1+\frac{J}{2\delta E_{z}}\right)^{2}-\left(1-\frac{J}{2\delta E_{z}}\right)^{2}}},
\end{align}
with integer $m$ and $n$ which fulfills Eq.~\eqref{eq:detRabi}. A comparison of the dynamics with and without synchronization is given in Fig.~\ref{fig:Synch}.

The second effect we observe is a phase accumulation for each individual spin during the microwave drive. During the CNOT gate a dynamic phase is acquired on the right (control) spin originating from the energy difference between the two blocks in Eq.~\eqref{eq:HamRWAblock}. While $\ket{\Psi_{R}}=\ket{\uparrow}$ states are oscillating with $\E^{-\frac{\I t c_{1}}{2}}$, $\ket{\Psi_{R}}=\ket{\downarrow}$ states oscillate with $\E^{-\frac{\I t c_{2}}{2}}$ which yields a relative phase after the ac spin flip, $\Phi_{R}\propto (c_{1}-c_{2})\tau\st{AC}$, on the right spin (see Eqs.~\eqref{eq:SubUp} and~\eqref{eq:SubDown}). Additionally, we observe a holonomic phase~\cite{Sjoeqvist2016} on the right spin which results from Rabi's equation for a full (half) cycle and can directly be seen in Eqs.~\eqref{eq:SubUp} and~\eqref{eq:SubDown} whether $\cos(\cdot)$ ($\sin(\cdot)$) becomes positive or negative depending on the choice of $m$ and $n$. We find the following analytic expressions for the phase difference on the right (control) spin after the ac spin flip,
\begin{align}
\Phi_{R}^{\text{AC}}=-\pi &\left[ \left(m-n+\frac{1}{2}\right)\phantom{\frac{2m+1}{a_{n,m}\left(1+\frac{J}{2\delta E_{z}}\right)}}\right.\nnb
&\left.-\frac{2m+1}{a_{n,m}\left(1+\frac{J}{2\delta E_{z}}\right)}\left(\delta E_{z}+\frac{J^{2}}{\delta E_{z}}-\frac{J}{2}\right)\right].
\label{eq:ACphaseR}
\end{align}
In order to find the phase accumulated during the full CNOT gate, consisting of dc pulse and ac pulse, the ac phase error, Eq.~\eqref{eq:ACphaseR} and the dc phase error, Eqs.~\eqref{eq:DCphaseL}~and~\eqref{eq:DCphaseR}, have to be combined. Considering the rotating frames for the dc and ac phase accumulation, we find the following results in the rotating frame of each individual spin, $R=\exp\left[-\I t(\omega_{1}S_{z,L}+\omega_{2}S_{z,R})/\hbar\right]$, with $\omega_1 = (E_z + \delta E_z/2)/\hbar$ and $\omega_2 = (E_z - \delta E_z/2)/\hbar$  (motivated by experiment~\cite{Zajac2017});
\begin{align}
\Phi_{L}&=\Phi_{L}^{\text{DC}}+\frac{\delta E_{z}}{2}\tau\st{DC}+\frac{\sqrt{\delta E_{z}^{2}+J^{2}}}{2}\tau\st{AC},\\
\Phi_{R}&=\Phi_{R}^{\text{DC}}-\frac{\delta E_{z}}{2}\tau\st{DC}+\Phi_{R}^{\text{AC}}-\frac{\sqrt{\delta E_{z}^{2}+J^{2}}}{2}\tau\st{AC}.
\end{align} 
This additional phase can either be compensated by adjusting $J$ such that the phase is a multiple of $2\pi$ (not possible in our regime of operation) or including an additional $z$-rotation directly after the CNOT gate with angles $\Phi_{l}=2k_{1}\pi -\Phi_{L}$ and $\Phi_{r}=2k_{2}\pi -\Phi_{R}$ with integers $k_{1,2}$. Simulations where we numerically integrate the time-dependent Schr\"odinger equation $i\hbar\dot{\Psi}(t)=\widetilde{H}(t)\Psi(t)$ support our analysis (see Fig.~\ref{fig:phasecorrection}). The highest fidelity can indeed be found after correcting the described phase shift. At this point it is worth mentioning that $z$-rotations in the experiment in Ref.~\cite{Zajac2017} and similar experiments ~\cite{Yoneda2017,Watson2017} can be performed by modifying the reference phase for the individual spins. This can be done rapidly and accurately in software with no additional microwave control required. 

\begin{figure}[t]
\begin{center}
\includegraphics[width=1.\columnwidth]{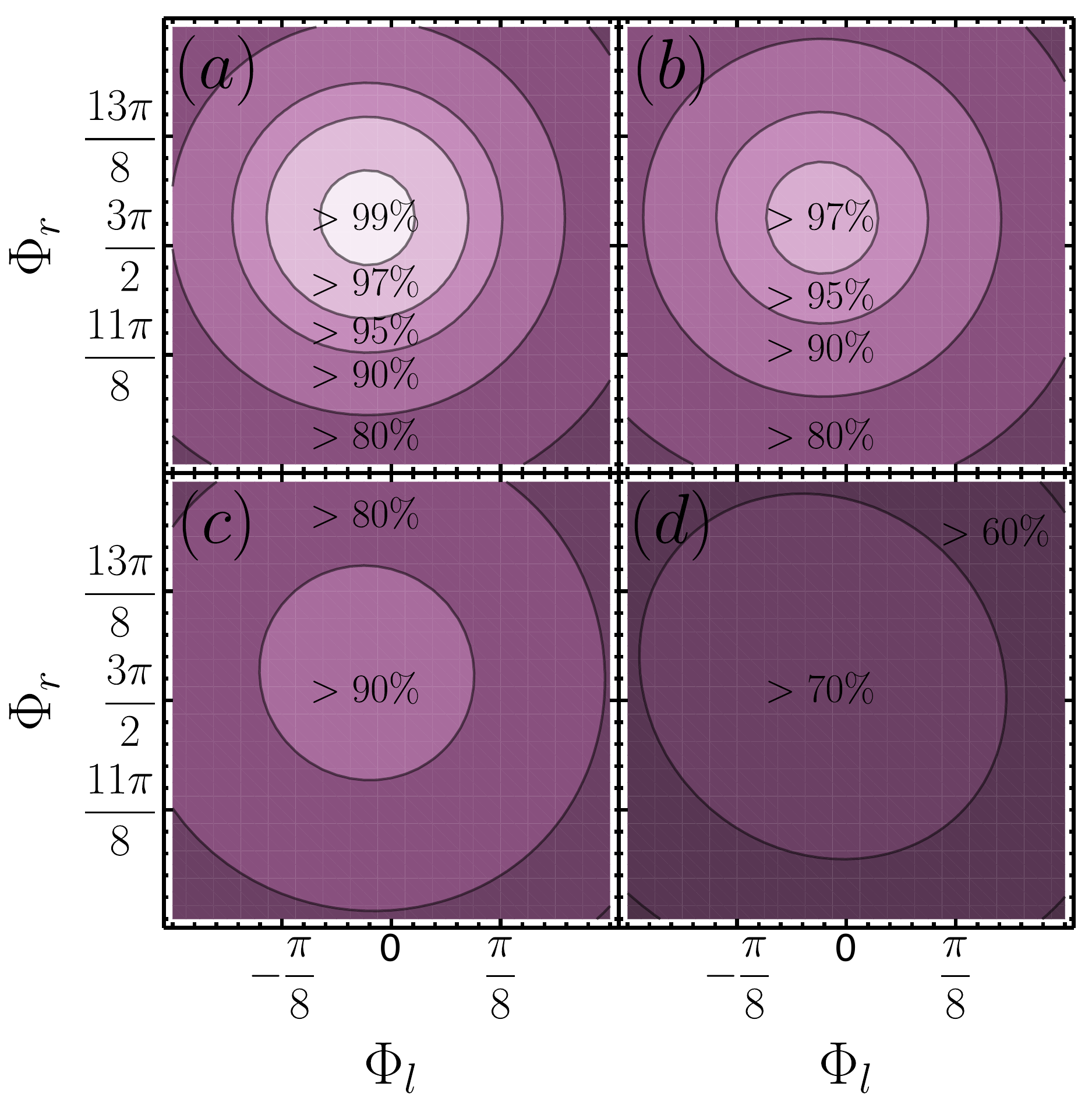}
\caption{Zoom in of the simulation in Fig.~\ref{fig:phasecorrection} in the presence of charge noise fluctuations with strength (a) $\sigma_{\delta J}=\unit[0]{MHz}$, (b) $\sigma_{\delta J}=\unit[0.33]{MHz}$, (c) $\sigma_{\delta J}=\unit[0.67]{MHz}$, and (d) $\sigma_{\delta J}=\unit[1.33]{MHz}$. For the simulation the fluctuations are assumed to be quasistatic and Gaussian distributed with standard deviation $\sigma_{\delta J}$ and mean $\langle \delta J \rangle\approx0$.}
\label{fig:noise}
\end{center}
\end{figure}

\subsection{Charge noise analysis}
\label{ssec:ACnoise}

In semiconductor devices charge noise is omnipresent\cite{Paladino2014}. In the simplest model, charge noise can be described as fluctuations of the electric potentials near the dot. Thus, charge noise couples to the two-qubit systems mainly through the exchange interactions due to its dependence on the detuning, tunneling, and confinement of the spins~\cite{Burkard1999,DasSarma2011}. To be precise, charge noise couples also to single spins through the same mechanism that allows EDSR to rotate the spin though fluctuations of the electron positions. This effect, however, is small as evidenced by Ref.~\cite{Yoneda2017}, thus will be neglected in the analysis below. 

In lowest order $J\rightarrow J+\delta J$ where $\delta J$ are fluctuations of the exchange energy due to charge noise, we find the following first-order corrections to the diagonal Hamiltonian~\eqref{eq:eigenDC1}-\eqref{eq:eigenDC4} in the adiabatic basis $\lbrace \ket{\uparrow\uparrow},\ket{\widetilde{\downarrow\uparrow}},\ket{\widetilde{\uparrow\downarrow}},\ket{\downarrow\downarrow}\rbrace$
\begin{align}
H\st{noise}=\frac{2J\delta J}{\delta E_{z}}(\widetilde{S}_{z}^{L}+\widetilde{S}_{z}^{R})-2\delta J \widetilde{S}_{z}^{L}\widetilde{S}_{z}^{R}.
\label{eq:chargenoise}
\end{align}
The first term induces single-qubit dephasing but is highly suppressed in the case of $J\ll\delta E_{z}$ since it has strength $\propto J/\delta E_{z}$. Therefore, large magnetic field gradients are beneficial for operating the two-qubit gate.
The second term couples longitudinally to the two-qubit gate operation since it has the same form as the dc pulse, $\propto S_{z}^{L}S_{z}^{R}$, and reduces the fidelity of the resulting two-qubit gate that only depends on the bare charge noise fluctuations $\delta J$. In experiments~\cite{Watson2017}, this is the limiting factor for the gate fidelity, since simple echo protocols which also filter out the desired two-qubit interaction would not work. Simulations assuming quasistatic noise show that for $\sigma_{\delta J}\equiv\sqrt{\langle\delta J^{2}\rangle-\langle \delta J\rangle^{2}}=\unit[0.33]{MHz}$ two-qubit gate fidelities $>97\%$ are still possible (see Fig.~\ref{fig:noise}~(b)). However, fluctuations twice (four times) as large already limit the gate fidelity to about $\approx 93\%$ ($\approx 77\%$) (see Fig.~\ref{fig:noise}~(c)~and~(d)), problematic for longer quantum algorithms. Mitigation of these effects is still possible through advanced pulse shaping or composite pulse sequences~\cite{Vandersypen2005}, complex dynamical decoupling sequences~\cite{De2014}, and reduction of the amplitude of the fluctuations, i.e., operating at a charge noise sweet spot. Also, a partial recovery of the fidelity is still achieved as the conditional spin flip during the frequency selective CNOT gate serves as a simple spin-echo sequence, decoupling the left spin and low-frequency charge noise if $\ket{\Psi_{R}}=\ket{\uparrow}$. This can be seen by approximating the CNOT gate as follows~\cite{Vandersypen2005};
\begin{align}
U\st{CNOT}&=\E^{-\I (H\st{rf}+H_{J})\tau\st{CNOT}/\hbar}\\ 
&\approx \E^{-\I H_{J}\,\tau/\hbar}\,\, \underbrace{\E^{-\I H\st{rf}\,\tau/\hbar}}_{\text{CNOT}}\,\, \E^{-\I H_{J}\,\tau/\hbar}
\label{eq:noiseDec}
\end{align}
with $\tau$ yet to be chosen. Here, the Hamiltonian $H\st{rf}$ only contains ac driving and the Hamiltonian $H_{J}$ contains the dc exchange interaction, thus, fluctuations due to charge noise. Assuming an ideal CNOT gate, only entries in Eq.~\eqref{eq:noiseDec} corresponding to $\ket{\Psi_{R}}=\ket{\downarrow}$ contain $J +\delta J$, therefore, are affected by charge noise.
Mathematically speaking, the time evolution $U\st{CNOT}$ only affects density matrix elements corresponding to $\ket{\Psi_{R}}=\ket{\downarrow}$ states which dephase with characteristic time $T^{-1}_{\phi}\approx\left\langle \delta J(t)^{2} \right\rangle$, while density matrix elements only consisting of $\ket{\Psi_{R}}=\ket{\uparrow}$ states are protected. In average, this will lead to a reduced influence of noise. Hypothetically, the larger variety of two-qubit quantum gates (modulating different transitions) and the partial intrinsic spin-echo can be used to construct more efficient charge noise decoupling sequences.

\section{Conclusion}
\label{sec:con}

In this paper, we have presented high-fidelity implementations of a dc-pulsed CPHASE gate and a single-shot two-qubit CNOT gate.

For the dc-pulsed CPHASE gate, we have provided a high-fidelity implementation using dc exchange pulses. We have analyzed two regimes for the exchange pulses, slow (adiabatic) and fast (instantaneous) exchange pulses, and have described how to compensate for residual spin-flip and phase errors. In the adiabatic regime, spin-flip errors are suppressed by the magnetic field difference and we have identified the phases which the individual spins accumulate during the two-qubit operation. By intersecting the CPHASE gate by single-qubit spin-flips to form a spin-echo sequence, spin-flip errors and local phases which the individual spins acquired during the CPHASE gate, can be avoided even for the non-adiabatic exchange pulses. 

For the ac single-shot two-qubit CNOT gate, we have presented a high-fidelity implementation through frequency-selective resonant modulations of the two-qubit transitions. By selecting different transition frequencies a larger set of two-qubit quantum gates is accessible allowing for more efficient algorithms. We are able to compensate all intrinsic errors due to off-resonant transitions by fine-tuning the ac driving amplitude such that the resonant and the off-resonant oscillations are synchronized. Additionally, we have identified phases which the individual spins accumulate during the two-qubit operation. These phases can be compensated for by performing single-qubit $z$-rotations after each CNOT gate. Our two-qubit gate implementation also incorporates a reduction of charge noise by suppression through large magnetic field gradients and a partial intrinsic spin-echo decoupling sequence. Using the synchronization technique and the analytic values of the accumulated phases, existing experiments are able to reach higher two-qubit gate fidelities exceeding $97\%$ under realistic assumptions. This opens the path to large-scale quantum computation previously limited by low-fidelity two-qubit gates. 

\acknowledgments
Research was sponsored by Army Research Office grant W911NF-15-1-0149, the Gordon and Betty Moore Foundation's EPiQS Initiative through grant GBMF4535, and NSF grant DMR-1409556. This research was partially supported by NSF through the Princeton Center for Complex Materials, a Materials Research Science and Engineering Center DMR-1420541.

\bibliography{lit6}

\end{document}